\documentclass{aastex}
\usepackage{emulateapj5}
\shorttitle{The origin of runaway stars}
\shortauthors{R.\ Hoogerwerf, J.H.J.\ de Bruijne \& P.T.\ de Zeeuw}
\begin{document}
%
%
\title{The origin of runaway stars}
\author{R.\ Hoogerwerf, J.H.J.\ de Bruijne \& P.T.\ de Zeeuw}
\affil{Leiden Observatory}
\affil{PO Box 9513, 2300 RA Leiden, The Netherlands}
\email{tim@strw.leidenuniv.nl}
%
%
\begin{abstract}
Milli-arcsecond astrometry provided by Hipparcos and by radio
observations makes it possible to retrace the orbits of some of the
nearest runaway stars and pulsars to determine their site of origin.
The orbits of the runaways AE~Aurigae and $\mu$~Columbae and of the
eccentric binary $\iota$~Orionis intersect each other $\sim$2.5~Myr
ago in the nascent Trapezium cluster, confirming that these runaways
were formed in a binary-binary encounter.  The path of the runaway
star $\zeta$~Ophiuchi intersects that of the nearby pulsar PSR
J1932$+$1059, $\sim$1~Myr ago, in the young stellar group Upper
Scorpius. We propose that this neutron star is the remnant of a
supernova that occurred in a binary system which also contained
$\zeta$~Oph, and deduce that the pulsar received a kick velocity of
$\sim 350$ km s$^{-1}$ in the explosion. These two cases provide the
first specific kinematic evidence that both mechanisms proposed for
the production of runaway stars, the dynamical ejection scenario and
the binary-supernova scenario, operate in nature.
\end{abstract}
%
%
\keywords{Astrometry --- stars: early-type --- stars: individual
($\zeta$~Oph, AE~Aur, $\mu$~Col, $\iota$~Ori) --- pulsars: individual
(PSR J1932$+$1059)}
%
%
\section{Introduction}
Runaway stars \citep{bla1961} distinguish themselves from the normal
population of early-type stars (spectral type O and B) by their large
peculiar velocities (up to 200 km~s$^{-1}$) with respect to the mean
Galactic rotation and by their isolated locations.  About 30\% of the
O stars and 5--10\% of the B stars are runaways. Since early-type
stars are relatively young (a few Myr to $\sim$50~Myr old), the
distances traveled by the runaways are relatively small (a few hundred
pc to a few kpc) and it is therefore possible, in some cases, to
identify the parent cluster, i.e., the stellar group where the runaway
originated \citep[e.g.,][]{bla1993}.

Two mechanisms for the production of runaway stars remain viable, the
{\it binary-supernova scenario} \citep{zwi1957,bla1961} and the
{\it dynamical ejection scenario} \citep*{pra1967,gb1986}. In
the former the runaway receives its velocity after a supernova
explosion in a massive close binary; after the explosion the binary
(sometimes) dissociates and the secondary starts to move through space
with a velocity comparable to its pre-explosion orbital velocity.  In
the latter scenario the runaway gains its velocity through a dynamical
interaction with one or more other stars. The most efficient encounter
is that between two hard binaries which, in most cases, results in the
ejection of two runaway stars and one eccentric binary
\citep{hof1983,mik1983}.

Which of the two mechanisms dominates the creation of runaway stars
has been debated vigorously in the past. The modest accuracy and
incompleteness of previous data sets did not allow distinguishing
convincingly between the two mechanisms: both scenarios were
consistent with the statistical properties of the ensemble of runaway
stars. The availability of Hipparcos milli-arcsec (mas) astrometry for
the nearby stars (ESA 1997), and pulsar astrometry provided by VLBI
and timing measurements, stimulated us to revisit this issue, as it
allows calculation of the past orbits of these objects with sufficient
accuracy to establish their parent group, and the mechanism that
formed them. A past encounter between a runaway star and a pulsar
would provide compelling evidence for the binary-supernova scenario,
while an encounter of two (or more) runaway stars would strongly
suggest the dynamical ejection scenario.  In the course of this work,
we could identify the specific formation scenario for three nearby
runaways (the pair AE~Aur \& $\mu$~Col and $\zeta$~Oph) with near
certainty.

In this Letter we summarize the results for these objects, as together
they demonstrate that both mechanisms for the creation of runaways
stars operate in nature. Details of the orbit integrations, the
Galactic potential used, and the simulations performed to validate the
results, as well as a discussion of the other nearby runaway stars and
pulsars with accurate astrometry, are reported in \citet[HBZ]{hbz2000}. 
\section{A stellar encounter in Orion}\label{aeaur_mucol}
The Orion association (Ori~OB1) has three known runaway stars: AE~Aur,
$\mu$~Col, and 53~Ari \citep{bla1961}. The first two of these form a
pair; they have almost similar spectral types (O9.5V and O9.5V/B0),
and move in opposite directions at 100 km~s$^{-1}$ each, leaving
Ori~OB1 about 2.5~Myr ago. These similarities led \citet{bm1954} to
suggest that the two stars were formed in the same event.
\citet{gb1986} proposed that AE~Aur, $\mu$~Col, and the massive
eccentric binary $\iota$~Ori (O9III+B1III) are the result of a
binary-binary encounter which ejected the two runaways.  Both stars
have normal rotational velocities (25 and 111~km~s$^{-1}$ for AE~Aur
and $\mu$~Col, respectively) and AE~Aur has a normal He abundance.
The He abundance of $\mu$~Col is unknown. Thus neither runaway shows
signs of previous mass transfer in a close binary system, and this
most likely excludes the binary-supernova scenario as the origin of
the two runaways. Dynamical ejection is by far the favorite scenario
for the origin of the high velocities of AE~Aur and $\mu$~Col.

We supplemented the Hipparcos astrometry with the best available
radial velocities in order to retrace the orbits of AE~Aur, $\mu$~Col
and $\iota$~Ori. We performed $2.5\times 10^6$ orbit integrations to
sample the errors in the positions and velocities of the objects (see
HBZ). The integrations show that $\sim$2.5~Myr ago the three stars
were very close together: the distribution of minimum separations
obtained from the orbit integrations is consistent with the stars
being located in exactly the same position in space at the same time,
$2.5\pm0.2$ Myr ago. We therefore conclude that the two runaway stars
and the binary $\iota$~Ori must once have been part of the same
cluster, and were all ejected following a dynamical interaction.

\begin{figure*}[t]
  \begin{center}
  \includegraphics[angle=0.0, width=15cm, 
                   clip=true, keepaspectratio=true]
                   {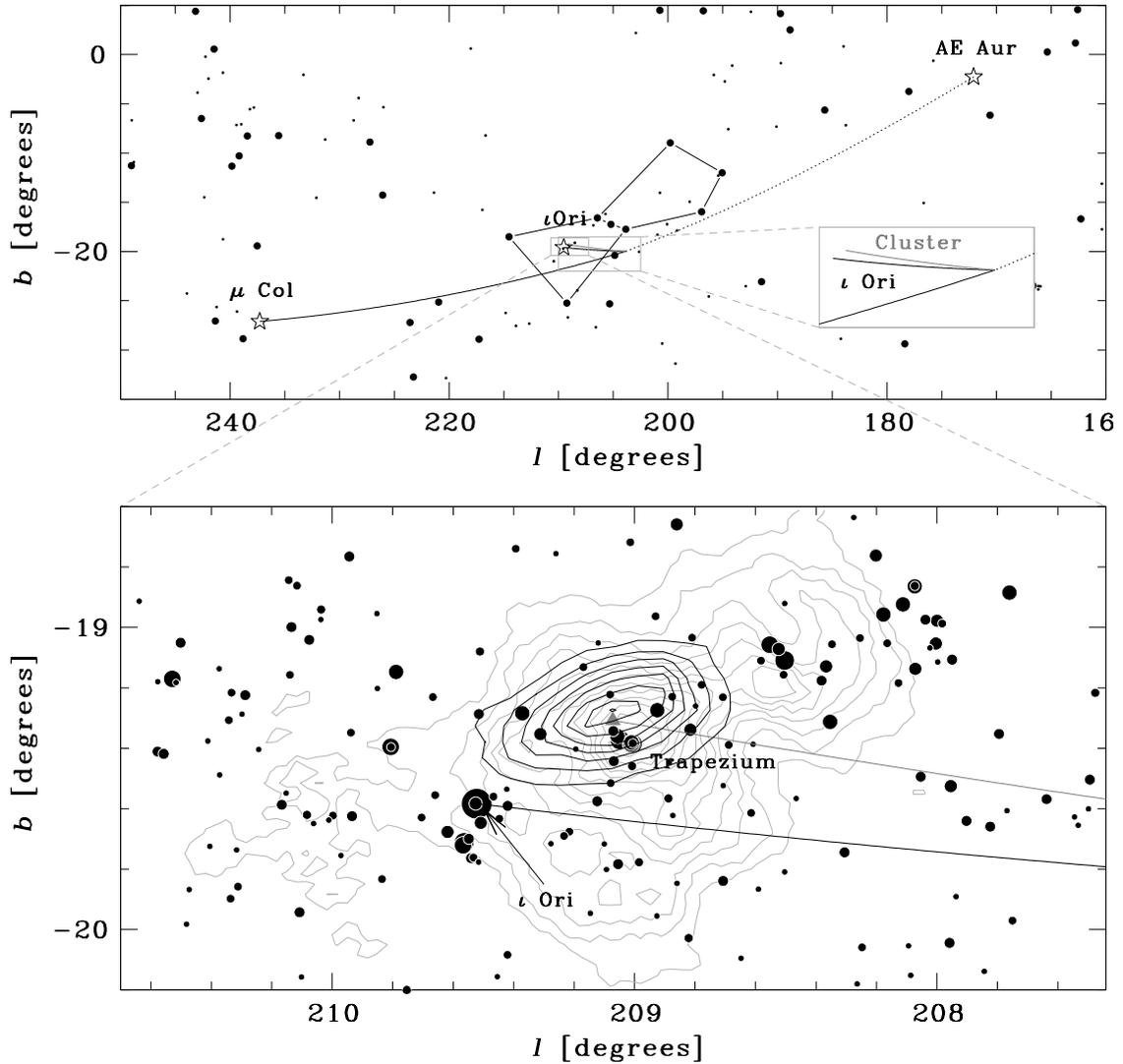}
  \end{center}
  \caption[]{{\it Top:\/} The past orbits on the sky of the runaways
AE~Aur (dotted line) and $\mu$~Col (solid line) and of the binary
$\iota$~Ori. The starred symbols depict the present position of the
three stars. The orbit of the parent cluster is denoted by the grey
solid line. The dots denote all stars in the Hipparcos Catalogue
brighter than $V = 3.5$~mag (large) and with $3.5~\mathrm{mag} \le V
\le 5$~mag (small, only the O- and B-stars) (cf.\ figure~1 in
\citealt{bm1954}). The Orion constellation is indicated for
reference. {\it Bottom:\/} The predicted position of the parent
cluster (black contours) together with all stars in the Tycho
Catalogue \citep{esa1997} in the field. The size of the symbols scales
with magnitude. The black and dark grey lines are the past orbits of
$\iota$~Ori and the Trapezium, respectively (see top panel).  The grey
contours display the IRAS 100 micron flux map, and mainly outline the
Orion Nebula.}
\label{fig:1}
\end{figure*}

A natural question to ask is which cluster hosted the four stars
before their encounter? Assuming that the center-of-mass motion of the
four stars was similar to that of the parent cluster, and using
conservation of linear momentum, provides the position and velocity of
the parent cluster $\sim$2.5~Myr ago. We integrated its orbit forward
in time to the present (Figure~\ref{fig:1}). The resulting properties
of the parent cluster, in particular its distance and position on the
sky, agree very well with that of the Trapezium cluster (see
Table~\ref{tab:1} and Figure~\ref{fig:1}).

Several other properties of the Trapezium cluster strengthen the
conclusion that this is the most likely candidate for being the parent
of AE~Aur, $\mu$~Col, and $\iota$~Ori. First, the Trapezium is a very
young cluster, $\sim$2~Myr \citep{ps1999}; its density ($>$20,000
stars pc$^{-3}$ in the center) is still high enough for stellar
encounters to occur. At the same time, the Trapezium is old enough to
have existed when the runaways left the cluster $\sim$2.5~Myr
ago. Secondly, the Trapezium shows a strong mass segregation
\citep*{zmw1993,hh1998}; this concentration of massive stars increases
the probability for dynamical encounters between stars. Thirdly, the
binary fraction in the Trapezium is high, 60--100\%
\citep{pro1994,wei1999}. This also increases the chance of dynamical
encounters since binary-binary collisions are the most efficient.
\section{A supernova in Upper Scorpius}\label{zetaoph}
\citet{bla1952} identified the O9.5V star $\zeta$~Oph as a runaway
originating in the Sco~OB2 association. The star could either have
left the Upper Scorpius subgroup $\sim$1~Myr ago or the Upper
Centaurus Lupus subgroup $\sim$3~Myr ago. The intrinsic properties of
$\zeta$~Oph (observed rotational velocity $v_{\rm rot} \sin i =
350$~km~s$^{-1}$ and He abundance $Y=0.40$) indicate that the star
previously experienced mass transfer in a close binary system. As
$\zeta$ Oph is single at present \citep{gb1986}, the binary must have
dissociated sometime in the past. This led to the suggestion that
$\zeta$~Oph is a runaway created in a binary supernova
explosion. Finding a compact object formed in the same event would
prove that this assumption is correct.

\begin{figure*}[t]
  \includegraphics[angle=0.0, width=8.5cm, 
                   clip=true, keepaspectratio=true]
                   {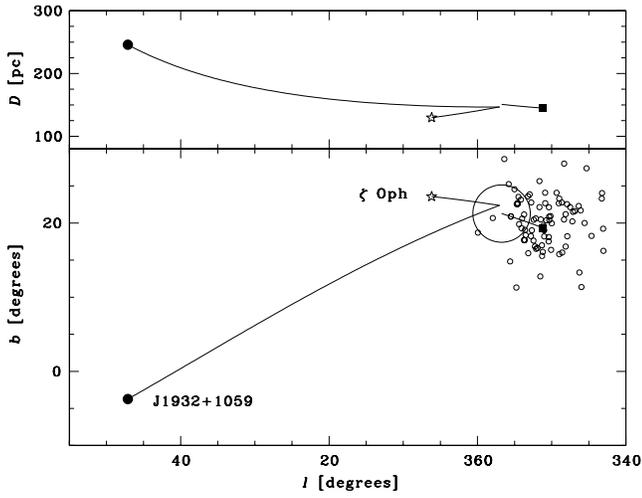}
  \caption[]{The orbits of $\zeta$~Oph, PSR~J1932$+$1059, and Upper
  Scorpius. The present positions are denoted by a star for the
  runaway, a filled circle for the pulsar, and by a filled square for
  the center of the association. The {\it top} panel shows the
  distance vs.\ Galactic longitude of the stars. The {\it bottom}
  panel shows the orbits projected on the sky in Galactic
  coordinates. The small open circles in the bottom panel denote the
  present-day positions of the O, B, and A-type members of Upper
  Scorpius \citep{zhb1999}. The large circle denotes the position of
  the association at the time of the supernova explosion (10~pc
  radius). This figure assumes a set of space motions consistent with
  all observables.}
\label{fig:2}
\end{figure*}

Radio pulsars are the only neutron stars for which reliable and
accurate proper motion measurements are available.  Of all pulsars in
the \citet*{tml1993} catalog, only seven are within 1~kpc, and have
proper motions with better than 10\% accuracy. Six of the seven cannot
be related to $\zeta$~Oph, based on their (three-dimensional) position
and velocity relative to $\zeta$~Oph, their position and velocity
relative to the Galactic plane, or their characteristic age
$P/(2\dot{P})$ (see HBZ for details). The pulsar that remains is PSR
J1932$+$1059 (a.k.a.\ B1929$+$10); it has a characteristic age of
$\sim$3~Myr, and it traversed the Upper Scorpius region about 1 Myr
ago if its (unknown) radial velocity $v_{\rm rad} =
200\pm50$~km~s$^{-1}$ (cf.\ Figure~\ref{fig:2}).

The pulsar currently moves away from the Galactic plane with a
$z$-velocity of $\sim$40~km~s$^{-1}$, and is located at a Galactic
latitude of $b \sim -4$ degrees. Assuming that most neutron stars are
created in or near the Galactic disk, PSR J1932$+$1059 must have
formed either recently or some 50~Myr ago, when its past orbit again
crossed the Galactic plane \citep[e.g.,][]{br1998}.  Taking into
account its characteristic age, it is natural to assume that the
pulsar was formed recently. The only site of active or recent star
formation along the pulsar's path is Upper Scorpius, and this young
stellar group therefore is the only likely birth-site for PSR
J1932$+$1059.

We integrated the orbits of $\zeta$ Oph, the pulsar, and Upper
Scorpius back in time in a standard Galactic potential. We performed
$3\times 10^6$ such integrations to sample the error distibutions in
position and velocity. The main uncertainties are the errors in the
parallax of $\zeta$ Oph ($7.1\pm0.7$ mas) and the pulsar ($5\pm1.5$
mas, \citealt{cam1995}), and the remaining range in $v_{\rm rad}$. The
distribution of the minimum distance between $\zeta$ Oph and the
pulsar is consistent with zero distance, i.e., with both objects being
in the same location, $1.0\pm0.1$ Myr ago in Upper Scorpius
(Figure~\ref{fig:2}).  This is strong evidence for the binary
supernova scenario.

The kinematics of the expanding HI shell which surrounds Upper
Scorpius requires a supernova explosion 1--2~Myr ago, and the
present-day mass function of the subgroup suggests that originally one
more massive star or binary ($\sim$40~$M_\odot$) must have been
present \citep{geu1992}. The simplest interpretation is that this was
a close binary containing $\zeta$~Oph and the progenitor of PSR
J1932$+$1059.

The value of $P/2(\dot P)$ is an uncertain age indicator, and the
$\sim3$ Myr for PSR J1932+1059 is consistent with the kinematicl age of 
1 Myr derived here. The implied period at birth is 0.18 seconds; the
current period is 0.22 seconds.

Pulsars are expected to receive a kick velocity ${\vec v}_{\rm kick}$
at birth \citep[e.g.,][]{lai2000}. Observations of the ensemble of
pulsars suggest that the magnitude of ${\vec v}_{\rm kick}$ is a few
hundred km s$^{-1}$ \citep{har1997,hp1997}.  Assuming that $\zeta$ Oph
and PSR J1932+1059 originated in the same binary allows an individual
determination of ${\vec v}_{\rm kick}$. Based on the magnitude of the
space velocities, and the angle between the orbits of $\zeta$ Oph and
the pulsar, we obtain $|{\vec v}_{\rm kick}| = 350\pm50$ km s$^{-1}$.
New VLBI observations of the proper motion and parallax of PSR
J1932+1059 are being obtained by Campbell (priv.\ comm.), and will
make it possible to improve this estimate further.

\section{Conclusions}\label{conclusions}
The cases described in the previous two sections provide the first
{\it specific} evidence that both the binary-supernova scenario and
the dynamical ejection scenario produce runaway stars. This result is
based on a detailed analysis of the orbits of three runaway stars, in
contrast to earlier investigations which mainly focused on the {\it
statistical} properties of a set of runaway stars.

Determination of the fraction of the runaway stars created by either
mechanism will put strong constraints on cluster formation theories
and on stellar evolution theories. Numerical simulations such as those
performed by \citet{por1999} show that runaway statistics can put
limits on, for example, the number of primordial binaries.  However,
the current samples of runaway stars and pulsars with accurate
astrometry are severely incomplete. The Hipparcos Catalog is complete
only for $V < 7.3-9$ mag (depending on latitude), and less than a
third of the O and B stars in it have a measured radial velocity.
Because of beamed radio emission, we cannot observe all pulsars, and
not all of those that do radiate in our direction have been found in
the Solar neighbourhood; of these, only a few have an accurately
measured proper motion. Even so, the available data supports the
tentative conclusion that both mechanisms contribute about equally to
the production of runaway stars (see HBZ).

The approach followed here promises many new results once the next
generation of astrometric satellites, e.g., FAME and in particular
GAIA, are launched and VLBI techniques are developed fully.  These
will improve the accuracy of the astrometry for stars and pulsars into
the micro-arcsecond regime. With a fainter limiting magnitude, many
more runaways will have well-determined positions and velocities. It
will then be possible to identify the parent clusters of these
objects, and to learn much about the ages and kick velocities of
individual pulsars.

\acknowledgements It is a pleasure to thank Adriaan Blaauw, Bob
Campbell, Nicolas Cretton, Ed van den Heuvel, Rob den Hollander,
Michael Perryman and Simon Portegies Zwart for useful comments and
suggestions.
\clearpage

\onecolumn
\begin{deluxetable}{lrrl}
\tablewidth{9truecm} \tabletypesize{\small} \tablecaption{Predicted
properties of the parent cluster of AE Aur and $\mu$ Col, and 
observed properties of the Trapezium
\label{tab:1}}
\tablehead{\colhead{Property}&  \colhead{Parent cluster}& \colhead{Trapezium}& \colhead{Unit}}
\startdata
$D$                             &              425--450&             450--500& pc\\
$(\alpha,\delta)$               &   $(83\fdg9,-5\fdg2)$&   $(83\fdg8,-5\fdg4)$& \\
$(\mu_{\alpha\ast},\mu_\delta)$ &          $(1.7,-0.2)$&          $(2.7,-0.9)$& mas~yr$^{-1}$\\
$(\ell,b)$                      & $(208\fdg9,-19\fdg2)$& $(209\fdg0,-19\fdg4)$& \\
$(\mu_{\ell\ast},\mu_b)$        &           $(0.9,1.4)$&           $(2.0,2.0)$& mas~yr$^{-1}$\\
$v_{\rm rad}$                   &                  27.6&                    24& km~s$^{-1}$\\
\enddata
\end{deluxetable}


%
\end{document}